\documentclass[epj]{svjour}
\usepackage{epsfig}

\newcommand{\he}{$^4$He$(\vec e, e' \vec p)^3$H{ }}
\newcommand{\h}{$^1$H$(\vec e, e' \vec p)${ }}
\newcommand{\gev}{(GeV/c)$^2${ }}
\newcommand{\eep}{$(e,e^\prime p)$}

\begin{document}
\sloppy
\title{Medium Modification of the Proton Form Factor}
\author{Steffen Strauch for the Jefferson Lab E93-049 collaboration%
}                     
%
%
\institute{Department of Physics,
  The George Washington University,
  Washington, DC 20052, USA} %
\date{Received: date / Revised version: date}
%
\abstract{I argue that the double ratio of proton-recoil
polarization-transfer coefficients, $P'_x$ and $P'_z$, of the
quasielastic \he reaction with respect to the elastic \h reaction is
sensitive to possible medium modifications of the proton form factor
in $^4$He. Recent measurements at both Mainz and Jefferson Lab of this
double ratio at four-momentum transfers squared between between 0.4
\gev and 2.6 \gev are discussed. I show that the data challenge
state-of-the-art conventional meson-nucleon calculations, as these are
unable to describe the results. The data hint at the need to include
medium modifications of the proton form factor, as predicted by a
quark-meson-coupling model, in the calculations.  A recently approved
follow-up experiment at a $Q^2$ of 0.8 \gev and 1.3 \gev with
unprecedented precision will provide one of the most stringent tests
of the applicability of various calculations.
\PACS{%
 {25.30.Dh}{Inelastic electron scattering to specific states} \and
 {24.70.+s}{Polarization phenomena in reactions} \and
 {14.20.Dh}{Protons and neutrons} \and
 {13.40.Gp}{Electromagnetic form factors} \and
 {13.88.+e}{Polarization in interactions and scattering} \and
 {27.10.+h}{$A \le 5$}
} 
}
%
\maketitle
\section{Introduction}

The standard nuclear-physics model describes nuclei as clusters of
protons and neutrons held together by a strong force mediated by meson
exchange. The success of the nuclear shell model demonstrates that
this approximation has been highly effective. A recent example is the
very good description of the differential cross section and left-right
asymmetry in the $^{16}$O$(e,e'p)$ reaction \cite{gao00} with a
fully relativistic distorted-wave impulse approximation (RDWIA).

However, protons and neutrons are not the fundamental entities of the
underlying theory, quantum chromodynamics. At nuclear densities of
about 0.17 nucleons/fm$^3$, nucleon wave functions have significant
overlap. In the chiral limit, one expects nucleons to lose their
identity altogether and nuclei to make a transition to a quark-gluon
plasma. For that reason, one may expect that, for some observables,
the free-nucleon approximation is a highly uneconomical approach.  The
use of medium-modified nucleons as quasiparticles might be a better
choice. To demonstrate that the interpretation of the experimental
data is more efficient in terms of medium modifications of nucleon
form factors, it is required to have excellent control of the inherent
many-body effects, such as meson-exchange currents (MEC) and isobar
configurations (IC). In addition, when probing nuclear structure, one
has to deal with final-state interactions (FSI).

In unpolarized A\eep{} experiments involving light- and medium-mass
nuclei, deviations were observed in the longitudinal/transverse
character of the nuclear response compared with the free-proton case
\cite{steen,ulmer,reffay}. Below the two-nucleon-emission threshold,
these deviations were originally interpreted as changes in the nucleon
form factors within the nuclear medium.  However, strong-interaction
effects on the ejected proton (final-state interactions) later also
succeeded in explaining the observed effect \cite{cohen}.  Still,
tantalizing hints of medium effects remain for unpolarized
longitudinal/trans\-verse separations in the
$^4$He(e,e$^\prime$p)$^3$H reaction \cite{Ma89,Du93}.

A reaction which is believed to have minimal sensitivity to
reaction-mechanism effects is polarization transfer in quasielastic
nucleon knockout (see, {\it e.g.}, \cite{Laget94}). The
polarization-transfer observables are sensitive to the properties of the
nucleon in the nuclear medium, including possible modification of the
nucleon form factor and/or spinor.  This can be seen from free
electron-nucleon scattering, where the ratio of the electric to
magnetic Sachs form factors, $G_E$ and $G_M$, is given by \cite{Ar81}:
\begin{equation}
\frac{G_E}{G_M} = -\frac{P'_x}{P'_z} \cdot \frac{E_e + E_{e'}}{2m_p}
\tan (\theta_e /2).
\label{eq:free}
\end{equation}
Here, $P'_x$ and $P'_z$ are the transferred polarizations, transverse
and longitudinal to the proton momentum (see \cite{Dipc}). The beam
energy is $E_e$, the energy (angle) of the scattered electron is
$E_{e'}$ ($\theta_e$), and $m_p$ is the proton mass. This relation was
used to extract $G_E/G_M$ for the proton (most recently in
\cite{Jones00,Gayou01,Gayou02}).

In the following, the results of recent \he experiments to study the
proton knock-out process will be discussed. In these experiments
$^4$He was chosen as target because of its high density and relatively
simple structure, which allows for RDWIA and microscopic
calculations. The kinematics of these experiments, quasielastic
scattering at low missing momentum with symmetry about the
three-momentum-transfer direction, minimize conventional many-body
effects, as will be demonstrated below.

\section{Experiments}

The first $^4$He polarization-transfer measurement was performed at
the Mainz microtron (MAMI) at a four-momentum squared $Q^2 = 0.4$ \gev
\cite{mainz4he}. More recently these measurements were extended to
$Q^2$ = 0.5, 1.0, 1.6, and 2.6 \gev in experiment E93-049 at Jefferson
Lab Hall A \cite{e93-049,str02}. Since these experiments were designed to
detect differences between the in-medium polarizations compared to the
free values, both $^4$He and $^1$H targets were employed (due to
beam-time constraints, only $^4$He data were acquired at $Q^2=2.6$
(GeV/c)$^2$).  The induced polarization, $P_y$, was also measured in
the JLab experiment \cite{str02}.

In each experiment, two high-resolution spectrometers were used, one
to detect the scattered electron and one to detect the recoiling
proton. The latter was equipped with a focal-plane polarimeter
(FPP). The proton-recoil-polarization observables were extracted by
means of the maximum-likelihood technique, utilizing the azimuthal
distribution of protons scattered off the graphite analyzer in the
FPP.

The results of the measurements are expressed in terms of the
polarization-transfer double ratio, in which the helium polarization
ratio is normalized to the hydrogen polarization ratio measured in the
identical setting:
\begin{equation}
R = \frac{(P_x'/P_z')_{^4\rm He}}{(P_x'/P_z')_{^1\rm H}}.
\label{eq:rexp}
\end{equation}
As a cross-check, the hydrogen results were also used to extract the
free-proton form-factor ratio $G_E/G_M$, which was found to be in
excellent agreement with previous data \cite{Jones00,Gayou01}.  

Nearly all systematic uncertainties cancel in $R$: the polarization-transfer
observables are to first order independent of instrumental asymmetries
in the FPP, and their ratio is independent of the electron-beam
polarization and carbon analyzing power. Small systematic
uncertainties are predominantly due to uncertainties in the spin
transport through the proton spectrometer.

\section{Results}

\begin{figure}[htb!]
\centerline{\epsfig{file=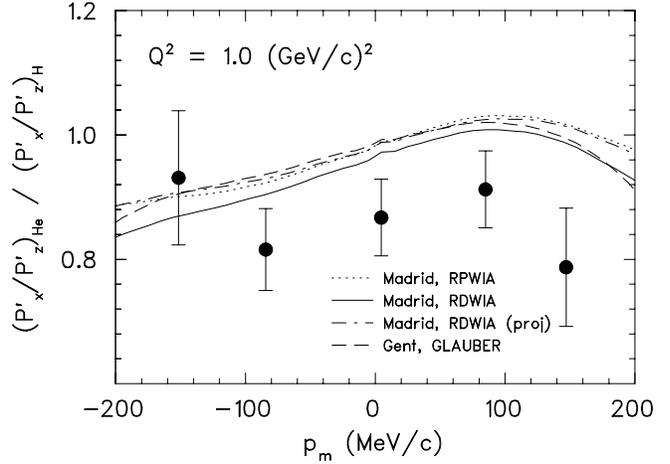,width=2.4in,angle=90}}
\caption{Polarization-transfer double ratio in \he at $Q^2$ = 1.0 \gev
  as a function of missing momentum. The data \protect\cite{str02} are
  compared to calculations by the Madrid \protect\cite{Ud98} and Gent
  \protect\cite{Ry99,debruynephd} groups. Error bars are
  statistical only; systematic uncertainties are much smaller.}
\label{fig:r_pm}
\end{figure}

Figure \ref{fig:r_pm} shows the results for $R$ at $Q^2$ = 1.0 \gev as
a function of missing momentum. The five data points correspond to
five bins in the acceptance within the single experimental
setting. Negative values of missing momentum correspond to the
recoiling nuclei having a momentum component antiparallel to the
direction of the three-momentum transfer.

The data are compared with several calculations, which were averaged
over the experimental acceptance. In the relativistic calculations by
the Madrid group \cite{Ud98}, the Coulomb gauge and the current
operator $cc1$ \cite{Fo83} were used.  The dotted line is a result of
a calculation in plane-wave impulse approximation (RPWIA).  This
calculation predicts $R\approx 1$ at $p_m=0$. It overpredicts the data
by about 10\%, although the missing-mass distribution of the data is
reasonably well described.  The relativistic distorted-wave impulse
approximation (RDWIA) gives a smaller value of $R$ ($\approx 3$\%)
than the RPWIA but still overpredicts the data (solid curve). The
calculation uses the MRW optical potential of \cite{McNeil83}. The
main difference in the results between this full calculation and the
RPWIA is due to the enhancement of the negative energy components of
the relativistic bound and scattered proton wave functions. This can
be demonstrated by projecting the wave functions over positive-energy
states (dash-dotted curve) resulting in essentially the same
polarization-transfer ratio as given by RPWIA. Also, a Glauber
calculation by the Gent group \cite{Ry99,debruynephd} (long-dashed
line) is unable to account for the data.

\begin{figure}[htb!]
\centerline{\epsfig{file=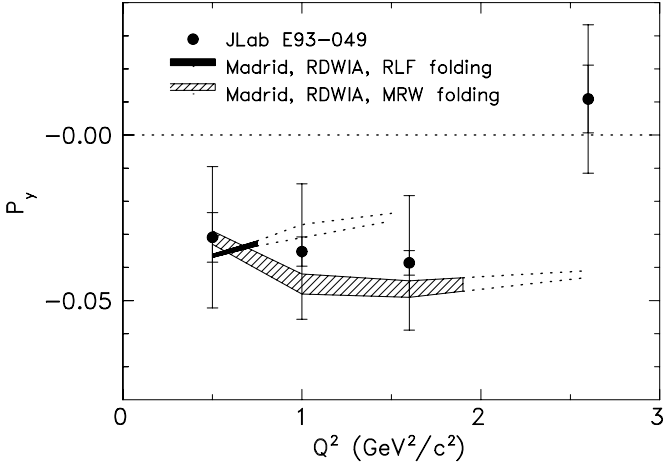,width=2.4in,angle=90}}
\caption{Measured values of the induced polarizations for the
{$^4$He$(e,e^\prime\vec{p})^3$H} reaction \cite{str02}. The inner
uncertainty is statistical only, the total uncertainty includes a
systematic uncertainty of $\pm$0.02, due to the imperfect knowledge of
the false asymmetries.  The hatched areas show the results for
different RDWIA calculations \protect\cite{Ud98} with the MRW
\protect\cite{McNeil83} and RLF \protect\cite{Horowitz85} relativistic
optical potentials.  All theoretical curves are averaged over the
experimental acceptance. Dotted curves indicate calculations in which
the optical potentials were extrapolated beyond the region of
validity.
\label{fig:induced}}
\end{figure}

The induced polarization $P_y$ in the $(e,e^\prime\vec{p})$ reaction
is identically zero, in the absence of FSI effects (in the
one-photon-exchange approximation). This observable thus constitutes a
stringent test of various FSI calculations. Figure~\ref{fig:induced}
shows the results, corrected for (small) false asymmetries, as a
function of $Q^2$.  One sees that the induced polarizations are small
for all measured $Q^2$ values.

The dashed and dot-dashed curves constitute RDWIA calculations by the
Madrid group \cite{Ud98} with the MRW \cite{McNeil83} and RLF
\cite{Horowitz85} relativistic optical potentials. For the
induced-polarization case, the RDWIA curves do not depend on the
nucleon form-factor model chosen in the calculations.
Figure~\ref{fig:induced} confirms the expected smallness of the
induced polarizations, and seems to indicate a reasonable agreement
with the RDWIA calculation.

However, due to the relatively large systematic uncertainties caused by
possible false asymmetries in the FPP, the test is not conclusive. A
far more detailed study of the induced polarization can be made with
data expected from JLab experiment E03-104 \cite{e03-104}.

\section{Discussion}

The disagreement between the data and the RDWIA calculations for the
polarization double ratio $R$ is puzzling, since, as mentioned
earlier, these relativistic calculations provide good descriptions of,
{\it e.g.}, the induced polarizations measured at Bates in the
$^{12}$C($e,e^\prime \vec p\,$) reaction \cite{woo98} and of
$A_{TL}$ in $^{16}$O($e,e^\prime p\,$) as previously measured at JLab
\cite{gao00}. To find a possible explanation for that disagreement, a
detailed study of model dependencies was performed.

\begin{figure}[htb!]
\centerline{\epsfig{file=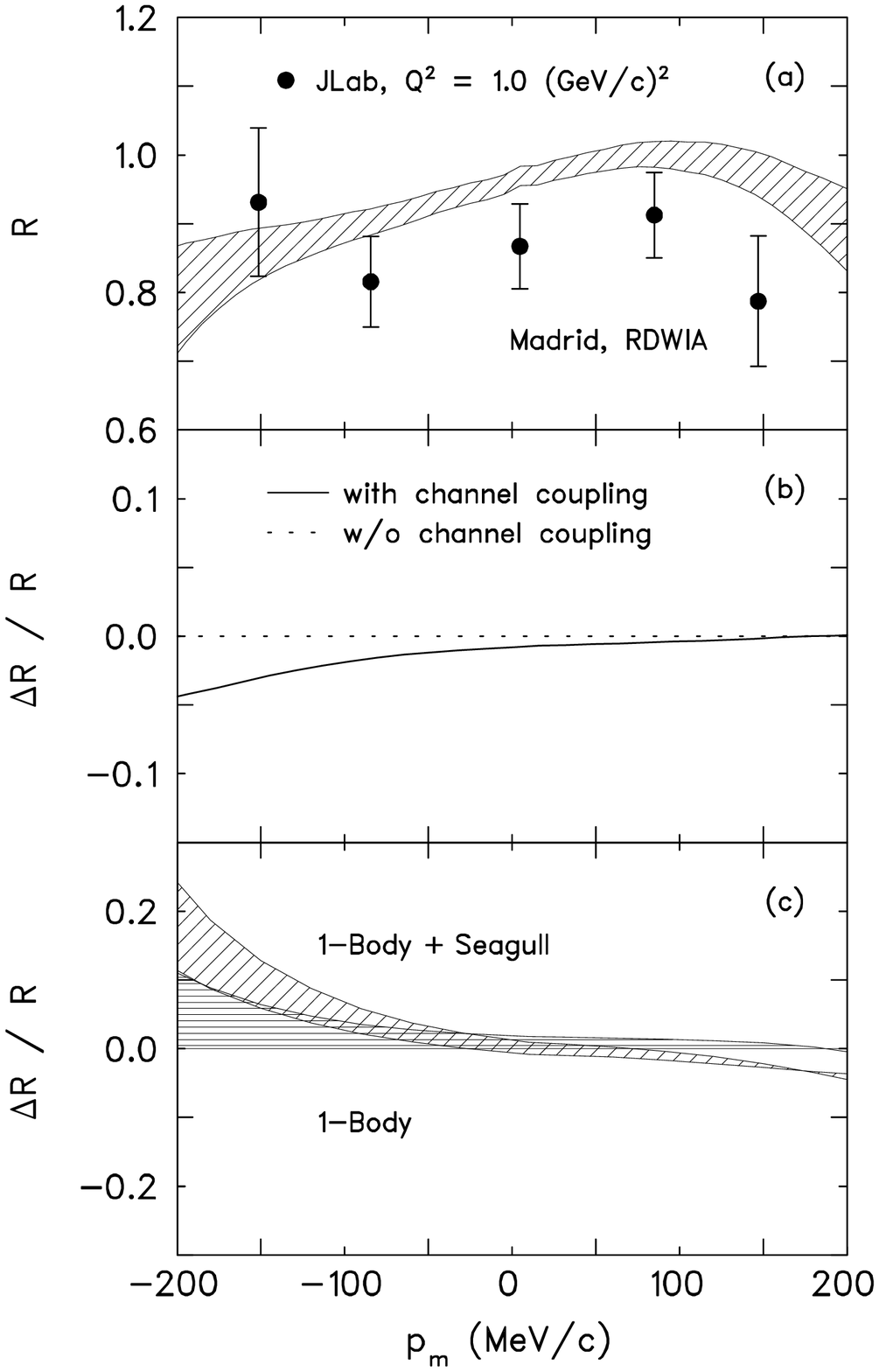,width=3.5in,angle=0}}
\caption{%
Study of model sensitivities to the polarization-transfer double ratio
in the \he reaction at $Q^2 = 1.0$ \gev.
(a) E93-049 polarization transfer double ratio at $Q^2 = 1.0$ (GeV/c)$^2$
along with a PWIA and different RDWIA calculations of the Madrid group
\protect\cite{Ud98}.
(b) Relative difference between the polarization transfer ratio
including channel coupling and without channel coupling. The
calculation follows the approach of \protect\cite{Kelly99}.
(c) Differences in the polarization double ratios relative to $R_{\rm
cc1}$ in the $^4$He$(\vec e,e' \vec p)^3$H reaction as a function of
missing momentum (perpendicular kinematics). Plotted are results for
different de Forest current operators with and without the MEC seagull
diagram. The 1-body $cc1$ result serves as a baseline. Calculation
courtesy of Meucci \cite{Meucci02}.  }
\label{fig:models}
\end{figure}

First, the model input in the calculations of the Madrid group was
studied. Figure \ref{fig:models}(a) shows again the $Q^2 = 1.0$ \gev
polarization-transfer double-ratio data. The hatched area indicates
the range of different RDWIA calculations. For these calculations,
various bound-state wave functions, current operators ($cc1$ and
$cc2$), and optical potentials were chosen. The optical potentials
include the MRW folding parameters \cite{McNeil83} and the RLF folding
parameters \cite{Horowitz85}. The latter are valid only up to $T_{\rm
lab} \approx 400$~MeV, and thus are already beyond its regime of
validity for $Q^2$ of 1.0 (GeV/c)$^2$. Different optical potentials
were constructed for each parameter set based upon the experimental
$^3$H density, based upon a simple Woods-Saxon density for $^3$H with
the same root-mean-square radius as the experimental one, and,
finally, one with an 30\% increased root-mean-square radius, resulting
in an unrealistic potential. Different parametrizations of the
free-nucleon form factors were also used. To a very good
approximation, the ratio $R$ is independent of such variations, if the
{\it same} parametrization is used in the $^4$He and $^1$H
calculations.  It is evident from Fig.~\ref{fig:models}(a) that there
is hardly any sensitivity in the polarization-transfer ratio at low
missing momentum (within the models examined) to the different and
partly extreme choices of optical potential, one-body current operator,
and bound-state wave function.

Next, the effect of channel coupling on $R$ for the \he reaction was
estimated. Recently Kelly \cite{Kelly99} investigated the sensitivity
of recoil-polarization observables in $A(\vec e,e'\vec p)B$ reactions
to channel coupling in final-state interactions (for $^{12}$C and
$^{16}$O). In these studies it was found that polarization-transfer
observables for proton knockout with modest missing momentum appear to
be quite insensitive to details of the final-state interaction,
including channel coupling.  Following the approach of Kelly
\cite{Kelly99}, the effect of channel coupling was estimated for the
\he{} reaction by calculating the relative difference between the
polarization transfer ratio with and without channel coupling.
Couplings between the proton $1s_{1/2}$ and neutron $1s_{1/2}$ states
were considered. Figure \ref{fig:models}(b) shows the results for a
four-momentum transfer of $Q^2 = 1.0$~(GeV/c)$^2$. The effect is
minimal, and on average is of the order of 1\% -- 2\%.

Finally, the effects of two-body interactions were considered.
Available model calculations indicate that these contributions are
smallest in quasi\-elas\-tic, parallel kinematics and at low missing
momentum; the kinematics of the experiments discussed here. In a
recent work, the Pavia group has studied meson-exchange currents in a
relativistic model for electromagnetic one-nucleon emission
\cite{Meucci02} (for $^{12}$C and $^{16}$O). Meucci has provided us
with similar calculations for $^4$He; these results are shown in
Fig.~\ref{fig:models}(c) for $Q^2 = 1.0$ \gev as relative differences
with respect to the one-body $cc1$ calculation.  The horizontally
hatched band covers the range of calculations with three different
one-body currents ($cc1$, $cc2$, $cc3$).  The calculations confirm
that, at low missing momentum, ambiguities due to the choice of the
1-body current ($cc1$, $cc2$, $cc3$) are small, of the order of
3\%. The inclusion of the two-body current in the form of the seagull
diagram with one-pion exchange (diagonally hatched band) has an
asymmetric effect on the polarization ratio about ${\mathbf p}_m = 0$;
the effect of MEC is also small. It reduces the polarization transfer
ratio by about 2\% on average.  In addition, these calculations
predict the effect of MEC to decrease with increasing four-momentum
transfer (not shown).

\begin{figure}[htb!]
\centerline{\epsfig{file=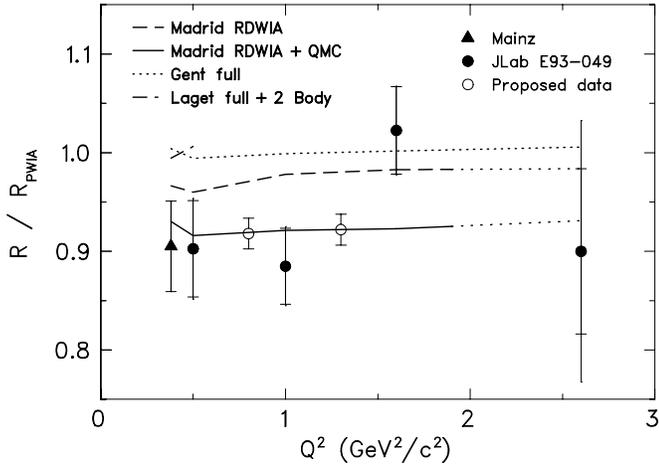,width=2.4in,angle=90}}
\caption{Superratio $R/R_{\rm PWIA}$ as a function of $Q^2$ (closed
circles). 
The dashed line shows the results of the RDWIA calculation of the
Madrid group \protect\cite{Ud98}. 
The dotted line shows the results of the full nonrelativistic model by
the Gent group \protect\cite{Ry99,debruynephd}.
The dash-dotted line shows the results of the full calculation,
including two-body currents, of Laget \protect\cite{Laget94}. 
The solid line indicates the full relativistic calculation of the
Madrid group \protect\cite{Ud98} including medium modifications as
predicted by a quark-meson-coupling model \protect\cite{Lu98}.  For
$Q^2 >$ 1.8 (GeV/c)$^2$ the Udias calculations maintain a constant
relativistic optical potential and are indicated as short-dashed
lines. The lines connecting the acceptance-averaged theory
calculations are to guide the eye only.  The open circles indicate the
projected data for experiment E03-104 \cite{e03-104}.
\label{fig:ratioplot}}
\end{figure}

Also the results of the microscopic calculations of Laget
\cite{Laget94} for $Q^2 = 0.4$ \gev and 0.5 \gev were found to be
nearly identical to the RPWIA results (see Fig.~\ref{fig:ratioplot})
indicating that reaction mechanisms like MEC, IC, or charge exchange
do not contribute significantly to $R$ in the present kinematics.


Does the present failure of state-of-the-art calculations to describe
$R$ indicates a breakdown of standard meson-nucleon calculations?
Probably not. It is likely that more complete and more complicated
many-body calculations eventually will account for the
data. Similarly, the observed cross-section scaling in deuteron
photodisintegration at high energies (see {\it e.g.} \cite{Gilman})
does not ``prove'' a breakdown of meson-nucleon calculations, but
rather indicates the onset of a regime where quark-gluon-inspired
models may give a more effective basis for the interpretation of the
data. In the same way, the observed reduction of $R$ in the \he
reaction motivates the use of a new approach in the interpretation of
the data.

Indeed, a calculation by Lu {\it et al.}~\cite{Lu98}, using a
quark-meson-coupling (QMC) model, suggests a measurable deviation from
the free-space form-factor ratio over the $Q^2$ range 0.0 $< Q^2 <$
2.5 (GeV/c)$^2$.  Note that the calculation is consistent with present
constraints on possible medium modifications for both the electric
form factor (from the Coulomb Sum Rule, with Q$^2$ $<$ 0.5 (GeV/c)$^2$
\cite{Jo95,Mo01,Ca02}), and the magnetic form factor (from a
$y$-scaling analysis \cite{Si88}, for Q$^2$ $>$ 1 (GeV/c)$^2$), and
limits on the scaling of nucleon magnetic moments in nuclei
\cite{Richter}. Similar measurable effects have been calculated in the
light-front-constituent quark model of Frank {\it et al.}~\cite{Fr96}.
Recently, Yakshiev {\it et al.}~\cite{Ya02} investigated possible
modifications to the electromagnetic form factors of the nucleons in
the $^4$He nucleus in the framework of a modified Skyrme model, up to
$Q^2$ = 0.6 (GeV/c)$^2$. Furthermore, the group of Miller is presently
studying medium effects in the framework of a chiral soliton model for
the proton \cite{MilPC}.

However, the notion of medium modification of single-particle
properties like, {\it e.g.}, the electromagnetic form factors of a
nucleon, in a nuclear environment is a purely theoretical concept
\cite{ArenPC}. Thus, distinguishing possible changes in the spatial
structure of nucleons embedded in a nucleus from more conventional
many-body effects is only possible within the context of a model;
presently the RDWIA calculation of the Madrid group.

\begin{figure}[htb!]
\centerline{\epsfig{file=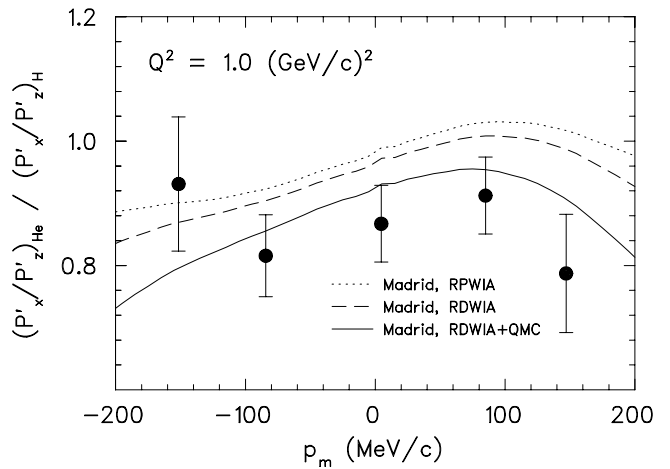,width=2.4in,angle=90}}
\caption{E93-049 polarization transfer double ratio at $Q^2 = 1.0$ (GeV/c)$^2$
along with a PWIA and different RDWIA calculations of Udias
\protect\cite{Ud98}.}
\label{fig:qmc}
\end{figure}

The RDWIA results are brought into excellent agreement with the data
by replacing the free-nucleon form factor with a density dependent,
medium-modified form factor based on the QMC model \cite{Lu98} (solid
curve in Fig.~\ref{fig:qmc}).

A summary of all available \he data \cite{mainz4he,str02} is shown in
Fig.~\ref{fig:ratioplot} along with different acceptance-averaged
calculations. The polarization-transfer double ratio is plotted as a
function of $Q^2$. In order not to obscure the result by small
kinematical differences between the individual $^1$H and $^4$He
measurements, data and calculations are shown with $R_{\rm PWIA}$ as a
baseline. The Mainz data point at $Q^2$ = 0.4 (GeV/c)$^2$ closely
coincides with the more recent results at $Q^2$ = 0.5 (GeV/c)$^2$ of
JLab.

On average, the RPWIA calculation (baseline) overestimates the data by
$\approx$ 10\%.  The dotted line shows the results of the full
nonrelativistic model of the Gent group \cite{Ry99,debruynephd}.  The
full calculation, including two-body currents, of Laget \cite{Laget94}
is only available for the lower two $Q^2$ values (dash-dotted).  Both
calculations are comparable to the results of RPWIA. The RDWIA
calculation (dashed curve) gives a slightly smaller value of $R$ but
still overpredicts the data.

The inclusion of medium-modified form factors as predicted by Lu {\it
et al.}  \cite{Lu98} into the RDWIA calculation gives a significantly
improved agreement with data, with the exception of the datum
at $Q^2 = 1.6$ \gev ($\chi^2$ per degree of freedom of 1.3 for the
five data points), in contrast to the use of free form factors
($\chi^2$ per degree of freedom of 2.2). All calculations by the
Madrid group \cite{Ud98} use the Coulomb gauge, the $cc1$ current
operator as defined in \cite{Fo83}, and the MRW optical potential of
\cite{McNeil83}.  The $cc2$ current operator gives higher values of
$R$, worsening agreement with the data. As discussed above, various
choices for, spinor distortions, current operators, and
relativistic corrections affect the theoretical predictions by $\le$
3\%, and presently cannot explain the disagreement between the data
and the RDWIA calculations.

Also shown in Fig.~\ref{fig:ratioplot} are the projected data points
at $Q^2 = 0.8$ \gev and 1.3 \gev for the newly approved JLab
experiment E03-104 \cite{e03-104}. This experiment will reduce the
statistical uncertainties in the double polarization ratio at each
$Q^2$ by over a factor of two compared to the previous measurements.
These two $Q^2$ values were selected since they lie in a region where
theoretical calculations are expected to be reliable. This measurement
will provide one of the most stringent tests to date of the
applicability of conventional meson-nucleon calculations.

Any medium modification of bound-nucleon form factors carries
implications for the nuclear EMC effect \cite{MilPC}. For example
strong constraints are placed on models of the nuclear EMC effect by
model-independent relations derived on the basis of quark-hadron
duality, which relate the medium modification of the electromagnetic
form factors to the modification of the deep-inelastic structure
function of a bound proton \cite{Mel02}.

\section{Summary}

Polarization transfer in the quasielastic \eep{ } reaction is
sensitive to possible medium modifications of the bound-nucleon form
factor, while at the same time largely insensitive to other reaction
mechanisms. The \he polarization-transfer double-ratio data are not
well described by modern relativistic RDWIA calculations using free
nucleons as quasiparticles. The data can be efficiently described if
medium-modified nucleons are used in the calculations. However, the
evidence from the present data is still limited. New high-precision
data expected from JLab Hall A should provide a more stringent test of
conventional meson-nucleon calculations.

This work was supported in part by the U.S.~Department of Energy under
grant DE--FG02--95ER40901. Southeastern Universities Research
Association (SURA) operates the Thomas Jefferson National Accelerator
Facility under U.S.~Department of Energy contract
DE--AC05--84ER40150.


\begin{thebibliography}{9}
\bibitem{gao00} J.~Gao {\it et al.}, Phys.~Rev.~Lett. {\bf 84}, 3265 (2000).
\bibitem{steen} G. van der Steenhoven {\it et al.},
  Phys.~Rev.~Lett. {\bf 57}, 182 (1986); {\bf 58}, 1727 (1987).
\bibitem{ulmer} P. Ulmer {\it et al.}, Phys.~Rev.~Lett. {\bf 59}, 2259 (1987).
\bibitem{reffay} D. Reffay-Pikeroen {\it et al.}, Phys.~Rev.~Lett. {\bf 60},
	776 (1988).
\bibitem{cohen} T.D. Cohen, J.W. Van Orden, and A. Picklesimer, 
	Phys.~Rev.~Lett. {\bf 59}, 1267 (1987).
\bibitem{Ma89} A. Magnon {\it et al.}, Phys.~Lett. {\bf B222}, 352 (1989). 
\bibitem{Du93} J.E. Ducret {\it et al.}, Nucl. Phys.~{\bf A553}, 697c (1993).
\bibitem{Laget94} J.M.~Laget, Nucl.~Phys.~A {\bf 579}, 333 (1994).
\bibitem{Ar81} A.I.~Akhiezer and M.P.~Rekalo, 
	Sov.~J.~Part.~Nucl. {\bf 3}, 277 (1974); 
	R.~Arnold, C.~Carlson, and F.~Gross, 
	Phys.~Rev.~C {\bf 23}, 363 (1981).
\bibitem{Dipc} With the initial and final electron momentum $\vec k_i$ and
$\vec k_f$, the coordinate system is given by the unit vectors
$\hat z = (\vec k_i - \vec k_f) / |\vec k_i - \vec k_f |$,
$\hat y = (\vec k_i \times \vec k_f) / |\vec k_i \times \vec k_f |$, and
$\hat x = \hat y \times \hat z$.
\bibitem{Jones00} M.K. Jones {\it et al.}, 
	Phys.~Rev.~Lett. {\bf 84}, 1389 (2000).
\bibitem{Gayou01} O.~Gayou {\it et al.}, 
	Phys.~Rev. C {\bf 64}, 038202 (2001).
\bibitem{Gayou02} O.~Gayou {\it et al.},
	Phys.~Rev.~Lett. {\bf 88}, 092301 (2002).
\bibitem{mainz4he} S.~Dieterich {\it et al.}, Phys.~Lett. {\bf B500}, 47 (2001).
\bibitem{e93-049} Jefferson Lab experiment E93-049 {\it Polarization
Transfer in the Reaction $^4$He$(e,e'p)^3$H in the Quasi-elastic
Scattering Region}, R.~Ent and P.~Ulmer, spokespersons.
\bibitem{str02} S.~Strauch {\it et al.}, 
        Phys.~Rev.~Lett. {\bf 91}, 052301 (2003).
\bibitem{Ud98} J.M.~Udias {\it et al.}, 
	Phys.~Rev.~Lett. {\bf 83}, 5451 (1991);
	J.A.~Caballero, T.W. Donnelly, E. Moya de Guerra, and
	J.M. Udias, Nucl.~Phys.~{\bf A632}, 323 (1998);
	J.M. Udias and J.R. Vignote, Phys.~Rev. C {\bf 62}, 034302 (2000).
\bibitem{Ry99} J. Ryckebusch, D. Debruyne, W. Van Nespen, and S. Janssen,
Phys.~Rev. C {\bf 60}, 034604 (1999).
\bibitem{debruynephd} D. Debruyne, Ph.D. thesis, University of Gent (2001).
\bibitem{Fo83} T.~de~Forest, Nucl.~Phys. {\bf A392}, 232 (1983).
\bibitem{McNeil83} J.A.~McNeil, L.~Ray, and S.J.~Wallace, Phys.~Rev.~C
{\bf 27}, 2123 (1983).
\bibitem{Horowitz85} C. J. Horowitz, Phys.~Rev.~C {\bf 31}, 1340 (1985).
\bibitem{e03-104} Jefferson Lab experiment E03-104 {\it Probing the
  Limits of the Standard Model of Nuclear Physics with the
  $^4$He$(e,e'p)^3$H Reaction}, R.~Ent, R.~Ransome, S.~Strauch, and
  P.~Ulmer, spokespersons.
\bibitem{woo98} R.J.~Woo {\it et al.}, Phys.~Rev.~Lett. {\bf 80}, 456 (1998).
\bibitem{Kelly99} J.J.~Kelly, Phys.~Rev.~C {\bf 59}, 3256 (1999).
\bibitem{Meucci02} A.~Meucci, C.~Giusti, and F.D.~Pacati, 
Phys.~Rev.~C {\bf 66}, 034610 (2002), and
A.~Meucci private communication.
\bibitem{Gilman} R.~Gilman and F.~Gross,  J.~Phys.~G {\bf 28}, R37 (2002).
\bibitem{Lu98} D.H.~Lu, K.~Tsushima, A.W.~Thomas, A.G.~Williams, and K.~Saito,
Phys.~Lett. {\bf B417} (1998) 217 and Phys.~Rev. C {\bf 60}, 068201 (1999).
\bibitem{Jo95} J. Jourdan, Phys.~Lett. {\bf B353}, 189 (1995).
\bibitem{Mo01} J. Morgenstern and Z.-E. Meziani, 
	Phys.~Lett. {\bf B515}, 269 (2001).
\bibitem{Ca02} J. Carlson, J. Jourdan, R. Schiavilla, and I. Sick,
        Phys.~Lett.~{\bf B553}, 191 (2003).
\bibitem{Si88} I. Sick, Comm. Nucl. Part. Phys. {\bf 18}, 109 (1988).
\bibitem{Richter} T.E.O.~Ericson and A.~Richter, 
	Phys.~Lett.~{\bf B183}, 249 (1987).
\bibitem{Fr96} M.R.~Frank, B.K.~Jennings, and G.A.~Miller, 
	Phys.~Rev. C {\bf 54}, 920 (1996).
\bibitem{Ya02} U.T.~Yakhshiev, U-G. Meissner, and A. Wirzba,
        Eur.~Phys.~J.~A {\bf 16}, 569 (2003).
\bibitem{MilPC} G.~Miller, private communication (2003).
\bibitem{ArenPC} H.~Arenh\"ovel, private communication (2003).
\bibitem{Mel02} W.~Melnitchouk, K.~Tsushima, and A.W.~Thomas, 
        Eur.~Phys.~J. A {\bf 14}, 105 (2002).
\end{thebibliography}
\end{document}